\documentclass[%
 reprint,
 amsmath,amssymb,
 aps,
]{revtex4-2}

\usepackage{graphicx}
\usepackage{dcolumn}
\usepackage{bm}
\usepackage{hyperref}
\usepackage{svg}

\begin{document}

\preprint{APS/123-QED}

\title{Enhanced dumbbell diffusion in a periodic potential by the elevator effect}

\author{B. A. Kiang}%
\affiliation{%
Institute Lorentz for Theoretical Physics, Leiden University, Leiden, the Netherlands
}%

\author{H. Schiessel}
\affiliation{
Cluster of Excellence, Physics of Life, TU Dresden, 01307 Dresden, Germany\\
Institut f\"ur Theoretische Physik, TU Dresden, 01062 Dresden, Germany
}%

\date{\today}

\begin{abstract}
We present molecular dynamics simulations of the random walk of a dumbbell---two beads connected by a spring---in a one-dimensional periodic potential and compare our results in limiting cases to theoretical analytical equations. The relevant parameters in this system are the spring constant, the equilibrium distance of the spring (relative to the periodicity of the potential), and the amplitude of the potential. Dumbbells with equilibrium distances incommensurate with the potential periodicity and with a sufficiently large spring constant exhibit enhanced diffusion. The diffusion constant can exceed that of a single bead in the same potential landscape. In this case, the dumbbell resembles a traction elevator, with the two connected beads acting as the elevator car and counterweight: the ``elevator effect''.
\end{abstract}

\maketitle


\section{Introduction}
Diffusion processes, i.e., the motion of particles navigating through energy landscapes under the combined influence of thermal fluctuations and deterministic forces, have been studied for a long time \cite{vankampen}. Here we investigate the diffusion of two beads connected by an ideal spring, known as a dumbbell, in a one-dimensional periodic potential. Whereas the diffusion of a single particle in periodic potentials has been solved analytically \cite{LJ,dieterich1977diffusion, Besselderiv}, the addition of an internal degree of freedom for the dumbbell leads to complex behavior. Understanding how the spring energy interplays with external potential landscapes can be interpreted as a prototype for the diffusion of composite objects through structured environments. The diffusion of such a dumbbell has potential applications across numerous fields: colloidal transport through arrays of potential wells created with holographic optical tweezers, which can be applied for continuously fractionating particles, biological cells, and macromolecules \cite{Lasertweezer3, Lasertweezer5, Lazertweezer10}; Gaussian chains in random potentials \cite{randompotential11}; dispersion of colloidal silica dumbbells, which can model crystal and glass formation of low-aspect ratio anisotropic particles and could have photonic applications and be used in electro-optical devices \cite{colloidalsilicaDB}; the so-called ”pom-pom polymer”, an idealized model for polymer melts with long-chain side branches and more than one junction point, modeled as a single backbone with multiple branches emerging from each end \cite{pompom}; a dumbbell model of enzymes, hydrodynamically coupled to their environment showing enhanced diffusion when bound to a substrate \cite{illien1, illien2}. Finally, the diffusion of dumbbells has been studied in other contexts: in a substrate of tilted potentials \cite{heinsalu, sancho2010rich, romero2004modelization, evstigneev2009interaction}, bucktooth potentials \cite{zhang2024directed}, on metal surfaces \cite{Ruckenstein, shea2011langevin}, and with a focus on the Kramers escape rate over a single potential barrier \cite{asfaw2012exploring}.

A singular aspect of diffusing dumbbells is their internal degrees of freedom, modeled by a spring that separates the two beads. This feature introduces a coupling between the relaxation dynamics of the spring and the external potential. Thus, the system is fundamentally different from single-particle diffusion. The coupling can bring about directional transport under driving forces \cite{directionaltransport}, a non-trivial dependence of transport coefficients on system parameters \cite{PhysRevE.91.022313} or, as we shall see, enhanced diffusion; these effects are relevant for transport in systems governed by internal flexibility or shape anisotropy, such as biological filaments or polymer chains.

For simplicity, we limit our analysis here to the Brownian motion of a dumbbell in a one-dimensional, spatially periodic potential $V$. The key parameters are the spring constant $k$ and equilibrium distance $x_{\mathrm{eq}}$ of the spring, and the barrier height $V_0$ and periodicity $a$ of the external potential. The Hamiltonian is then given by
\begin{equation}
H=\frac{m\dot{x}_1^2}{2}+\frac{m\dot{x}_2^2}{2}+V(x_1)+V(x_2)+\frac{k}{2}(x_2-x_1-x_\mathrm{eq})^2
\end{equation} 
where $x_1$ and $x_2$ denote the positions of the left and of the right beads.
The system is simulated by using the Langevin thermostat in molecular dynamics \cite{Langevintherm}. The equations of motion of the beads obey
\begin{align}
m\ddot{x_1}&=-\frac{d}{dx_1}V(x_1)+k(x_2-x_1-x_\mathrm{eq})-m\zeta\dot{x}_1+W_1(t),\\
m\ddot{x_2}&=-\frac{d}{dx_2}V(x_2)-k(x_2-x_1-x_\mathrm{eq})-m\zeta\dot{x}_2+W_2(t).
\end{align}
We set the physical parameters to $k_BT=m=\zeta=1$. Here $k_B$ is the Boltzmann constant, $T$ the temperature, $m$ the mass of the particle, and $\zeta$ its friction coefficient. $W(t)$ is a random force acting on the beads, uncorrelated in time and between the beads with variance $\langle W_i(t)W_j(t') \rangle=2k_BTm\zeta\delta_{ij}\delta(t-t')$.  We simulate dumbbell diffusion for different values of the spring constant $k$ and equilibrium length $x_\mathrm{eq}$ and for different barrier heights $V_0$ of the potential but keep the periodicity $a$ of the potential fixed at $a=16$. 

\section{Theory}
For a single particle in a spatially periodic, one-dimensional potential $V\left(x\right)$ ($x$: position of particle), the diffusion constant $D$ is given by the Lifson-Jackson formula \cite{LJ, Besselderiv} 
\begin{equation}
\label{DIFFUSION}
\frac{D}{D_0}=\frac{a^2}{\int_0^a dx\, \exp(V(x))\,\int_0^a dx\, \exp(-V(x))},
\end{equation}
where $V $ is measured in units of $k_B T$. $D_0$ denotes the diffusion in the absence of an external potential; here $D_0=k_B T/\zeta=1$. 
We use here the Lifson-Jackson formula that is derived for the overdamped case, even though our simulations start from the underdamped equations, because the Langevin thermostat describes thermalized motion including inertia, fluctuations and dissipation. The finite mass $m$ of the beads introduces a velocity relaxation time of $m/\zeta$ (equal to $1$ in our case). As the inertial time scale is much faster than the time scale required to cross energy barriers (here order of magnitude $10^2$), velocities are equilibrated rapidly and do not influence the observed motion. The Langevin dynamics and Brownian dynamics reach an equivalence in terms of the positions of the beads, and, regardless of whether the dynamics at short times are underdamped or overdamped, at long times the repeated energy barrier crossings will lead to an effective diffusive motion. Specifically, we assume here the potential
\begin{equation}
V(x)=\frac{V_0}{2}\cos\left(\frac{2\pi x}{a}\right),
\end{equation}
for which it follows from Eq.~\ref{DIFFUSION} that the diffusion constant obeys
\begin{equation}
\label{bessel}
\frac{D}{D_0}=\frac{1}{\left[I_0\left(V_0/2\right)\right]^2},
\end{equation}
where $I_0$ is the modified Bessel function of the first kind \cite{dieterich1977diffusion}.

As a first example, consider a dumbbell with a spring equilibrium length that coincides with the period of the potential, $x_\mathrm{eq}=a=16$. From Eq.~\ref{bessel} follow immediately two limits for the diffusion of its center of mass (CM), $x_\mathrm{CM}$, $x_\mathrm{CM}=\left(x_1+x_2\right)/2$. For the limit $k \rightarrow 0$, the two beads are decoupled and thus diffuse independently of each other, resulting in a mean-squared displacement of $x_\mathrm{CM}$ that is reduced by a factor $1/2$ compared to the diffusion of a single bead, Eq.~\ref{bessel}. In the opposite limit, $k \rightarrow \infty$, the two beads are forced to cross the top of the barriers together, so that the dumbbell experiences an effective potential with twice the amplitude. From this follows that the diffusion constant is given again by Eq.~\ref{bessel} but with $V_0$ replaced by $2V_0$ and with a factor $1/2$ in front of the Bessel function, as the dumbbell experiences twice the friction of a single bead. In summary, the limiting cases are

\begin{equation}
\label{bessel2}
\frac{D}{D_0}\rightarrow
\begin{cases}
    \frac{1}{2\left[I_0\left(V_0/2\right)\right]^2} & \text{for} \quad k \rightarrow 0,\\
    \frac{1}{2\left[I_0\left(V_0\right)\right]^2}  & \text{for} \quad k \rightarrow \infty.\\ 
\end{cases}
\end{equation}

\section{Simulation results}

\begin{figure}
\includegraphics[width=\linewidth]{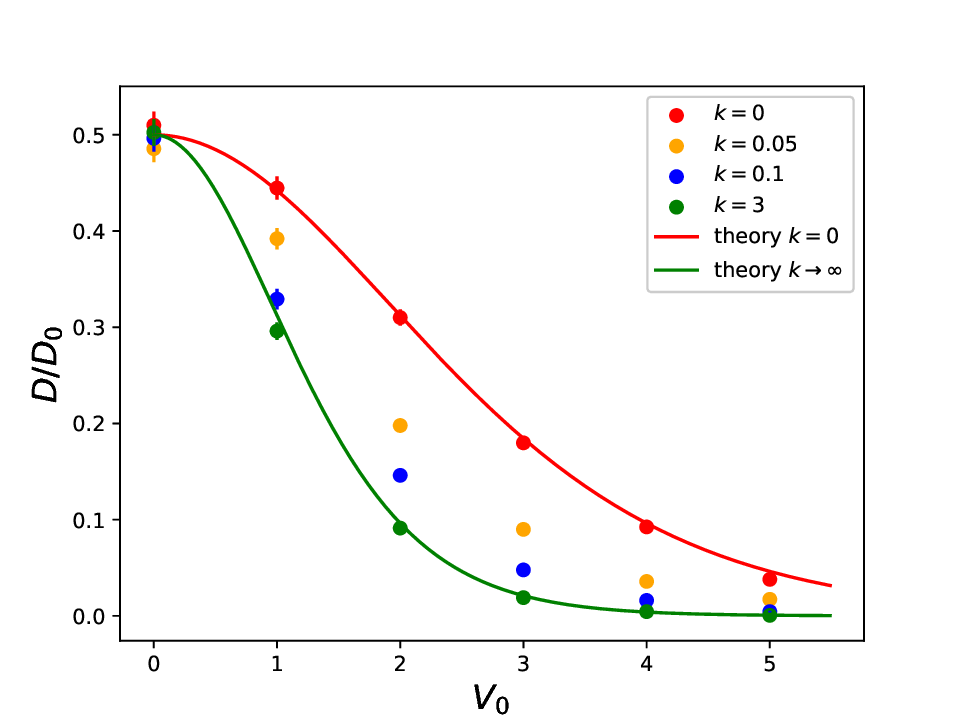}
\caption{\label{D,V} Diffusion constant $D$ as a function of barrier height $0\leq V_0\leq 5$, for different values of spring constant $k$ at spring equilibrium distance equal to the potential period, $x_\mathrm{eq}=a=16$. The solid lines are theoretical limits: the red line corresponds to two independent beads ($k=0$) and the green line to a rigid dumbbell ($k \rightarrow \infty$).}
\end{figure}

In Fig.~\ref{D,V}, we plot the diffusion constant of the dumbbell as a function of the barrier height for different values of the spring constant $k$. Here and in the following each data point results from $N=5\cdot 10^4$ MD simulation runs for $10^5$ time steps, where we allow the system to equilibrate for $2\cdot 10^3$ time steps. The time step is set to $\Delta t=0.005$. For each run, we determine the slope of the squared displacement after equilibration, and average over these slopes in order to obtain the diffusion constant. 

Here the equilibrium length $x_\mathrm{eq}$ of the spring is kept equal to the potential period $a=16$. The data are shown together with the two theoretical limits $k \rightarrow 0$ and $k \rightarrow \infty$ from Eq.~\ref{bessel2}. The MD simulations suggest that with increasing coupling the diffusion constant is reduced in a monotonic fashion, reflecting the fact that a ``jump'' of the dumbbell by one period of the potential gets increasingly costly. This reflects the transition from single-beads barrier crossing to two-beads barrier crossing due to stronger coupling of the beads. Also note that for the smallest non-vanishing spring constant simulated here, $k=0.05$, the diffusion is already substantially slowed down and that for the largest spring constant, $k=3$, the data points are already indistinguishable from the rigid system. That the approach to the $k \rightarrow \infty$-limit is indeed very rapid, can be observed in Fig.~\ref{D,k}, where we increase the spring constant in small steps from $k=0$ to $k=0.5$ for a potential barrier height of $V_0=2$.

\begin{figure}
\includegraphics[width=\linewidth]{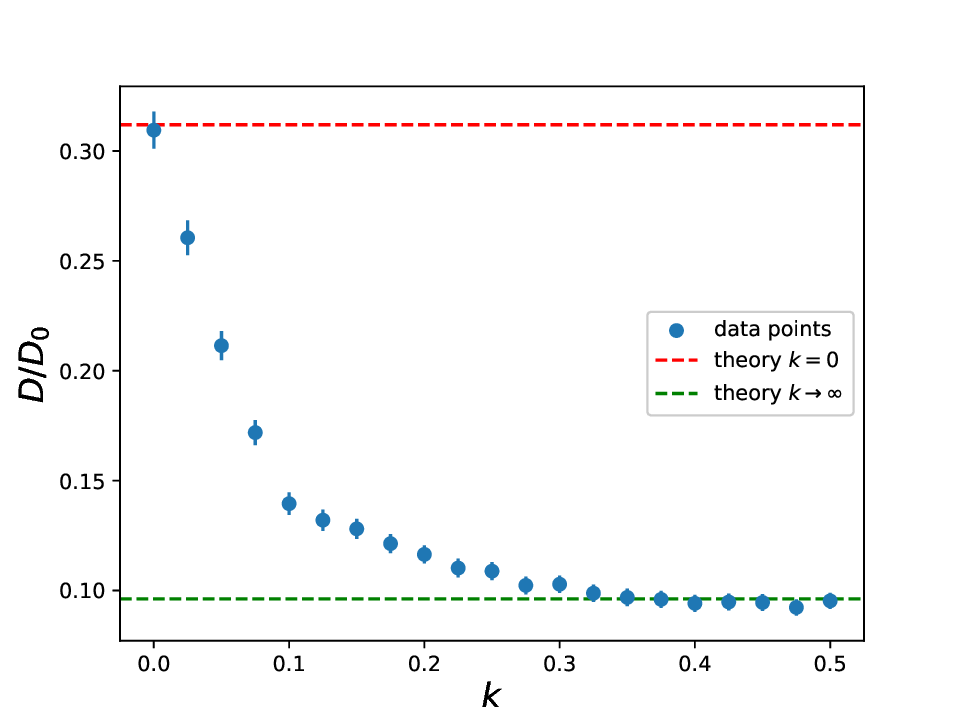}
\caption{\label{D,k} Diffusion constant $D$ as a function of spring constant $k$, for barrier height $V_0=2$ and equilibrium length $x_\mathrm{eq}=16$. The horizontal dashed lines indicate the asymptotic cases $k=0$ and $k\rightarrow \infty$.}
\end{figure}

\begin{figure}
\includegraphics[width=\linewidth]{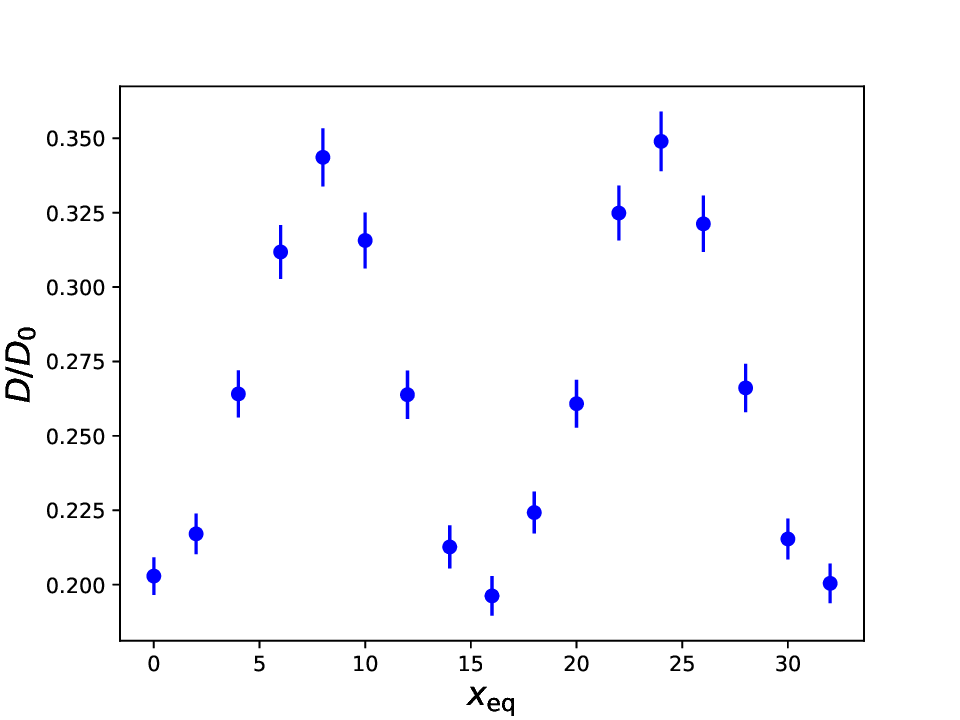}
\caption{\label{D,q1} Diffusion constant $D$ as a function of equilibrium length $0\leq x_\mathrm{eq}\leq 32$, for $V_0=2$ and $k=0.05$.}
\end{figure}

So far we considered the case that the equilibrium length of the spring, $x_\mathrm{eq}$, coincides with the period of the potential. In the following, we study how the diffusion is affected by the value of $x_\mathrm{eq}$. In Fig.~\ref{D,q1} we show the result from our MD simulations for $V_0=2$ and $k=0.05$ where we vary $x_\mathrm{eq}$ from $0$ to $32=2a$. We find a periodic dependence of the diffusion constant on $x_\mathrm{eq}$ with a periodicity $a$. 
Minima are found for equilibrium lengths that are commensurate with the periodicity of the potential, since in this case both beads can only avoid to cross barriers together if the spring gets deformed.

This can be best demonstrated for the strong spring limit, $k \rightarrow \infty$, that can be calculated  analytically.
Here the total potential energy of the dumbbell is given by
\begin{align}
V(x)&=\frac{V_0}{2}\left[\cos\left(\frac{2\pi x} {a}\right)+\cos\left(\frac{2\pi \left(x+x_\mathrm{eq}\right)} {a}\right)\right]\nonumber\\
&= V_0 \cos\left(\frac{\pi x_\mathrm{eq}} {a}\right)\cos\left(\frac{2\pi x} {a}+\frac{\pi x_\mathrm{eq}} {a}\right)
\end{align}
with $x=x_1$.
Plugging this potential into Eq.~\ref{DIFFUSION} and accounting for the fact that the friction of the dumbbell is twice that of a single bead we arrive at
\begin{equation}
\label{DIFFUSION2}
\frac{D}{D_0}=\frac{(2 \pi)^2}{\int_\alpha^{\alpha+2\pi}  e^{A\cos\theta}d\theta\,\int_\alpha^{\alpha+2\pi} e^{-A\cos\theta}d\theta}
\end{equation}
with $A=V_0\cos(\pi x_\mathrm{eq}/a)$ and $\alpha=\pi x_\mathrm{eq}/a$. Thus, we obtain for the stiff spring limit
\begin{equation}
\label{strong spr diff}
\frac{D}{D_0}=\frac{1}{2\left[I_0\left(V_0\cos\left(\pi x_\mathrm{eq}/a\right) \right)\right]^2}.
\end{equation}
For $x_\mathrm{eq}=a$, one recovers again Eq.~\ref{bessel2} (case $k \rightarrow \infty$). Fig.~\ref{D,q strong} presents a comparison between Eq.~\ref{strong spr diff} and MD simulations of a dumbbell with a strong spring, $k=2$, in a potential with $V_0=2$, showing excellent agreement between theory and simulations.

\begin{figure}
\centering
\includegraphics[width=\linewidth]{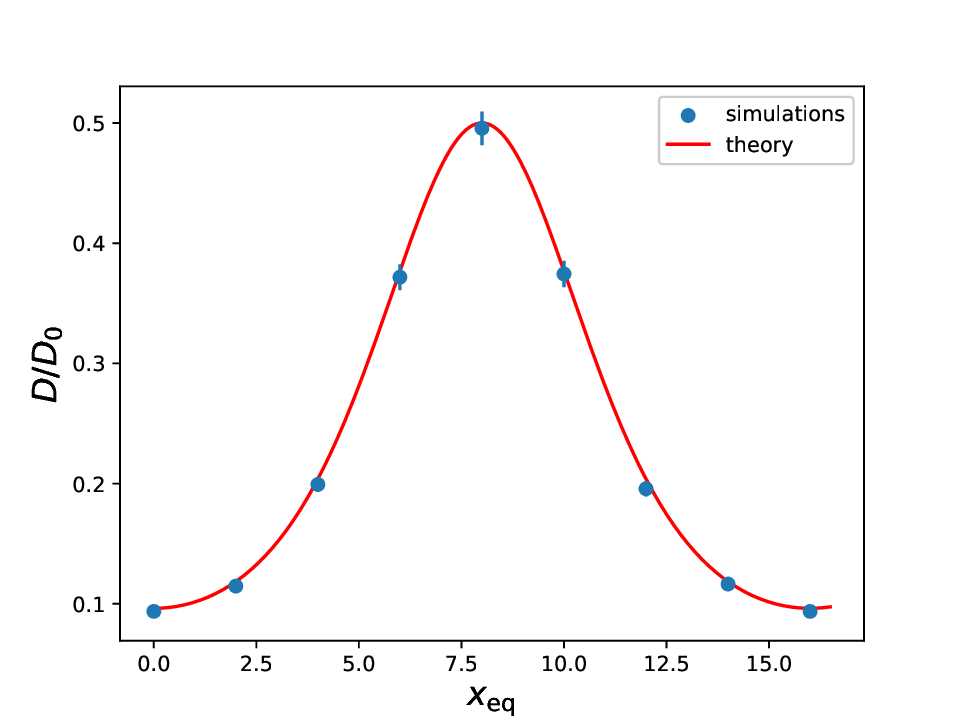}
\caption{\label{D,q strong}Simulation results for $D$ as a function of the spring equilibrium length $x_\mathrm{eq}$ for a strong spring $k=2$, compared to Eq.~\ref{strong spr diff}, the theoretical expectation for a rigid dumbbell (solid line).}
\end{figure}

\section{The ``elevator effect''}
In figs.~\ref{D,q1} and \ref{D,q strong} we observe that the diffusion constant features a maximum when the equilibrium length obeys $x_\mathrm{eq}=n a+1/2$ with $n$ denoting an integer. This agrees with the qualitative findings in Ref.~\cite{heinsalu}.
For a diffusing dumbbell with an undeformed spring this leads to a situation where a change in the potential energy of one bead is exactly compensated by that of the other bead. E.g., when one bead sits in a minimum position of the potential, the other bead is on top of a barrier. This effect explains why in this situation the diffusion is maximized. A similar effect is used for elevators, where the weight of the elevator cabin is (at least partially) compensated by a counterweight that moves in the opposite direction. Due to this analogy we call our observation of enhanced dumbbell diffusion the ``elevator effect''.

\begin{figure}
\includegraphics[width=\linewidth]{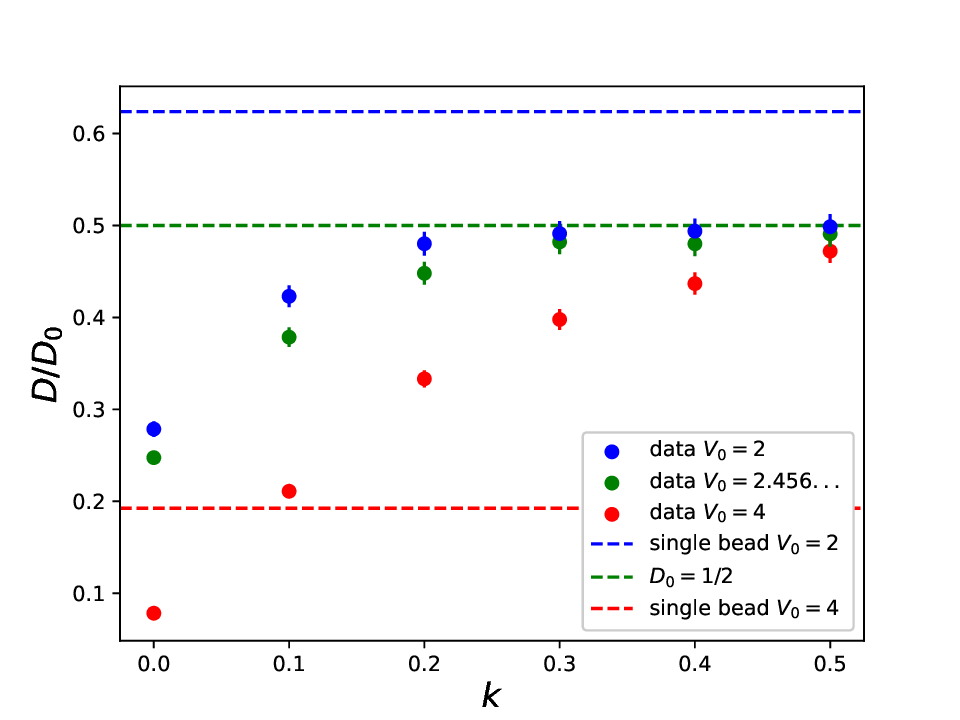}
\caption{\label{elevator 2} Diffusion constant of a dumbbell with $x_\mathrm{eq}=a/2$ in potentials with $V_0=2$ (blue), $V_0=2.456...$ (green) and $V_0=4$ (red), as a function of the spring constant $k$. Dashed lines indicate diffusion constants of single particle in the same potentials. Note that for $V_0=2.456...$ (green) the single bead diffusion constant equals $1/2$, the same value as for this dumbbell in the rigid spring limit, independent of $V_0$.}
\end{figure}

This raises the question of whether dumbbells can use the elevator effect to diffuse faster in an external periodic potential than individual beads, even assuming that the friction constant of a dumbbell is twice that of a single bead. The answer to this question can be found in Fig.~\ref{elevator 2} which plots the diffusion constant of the dumbbell with equilibrium length  $x_\mathrm{eq}=a/2=8$ as a function of $k$ for three different values of $V_0$. The horizontal dashed lines indicate the diffusion constants of single beads for the corresponding $V_0$-values. All three curves start at $k=0$ with a diffusion constant equal to half the diffusion constant of a single bead, and then increase with increasing spring strength until they finally asymptotically reach the value $D/D_0=1/2$.

For the smallest value of $V_0$, $V_0=2$, this stiff spring limit lies below the value of a single bead which is $D/D_0=1/\left(I_0(1)\right)^2=0.62...$, i.e., the stiff dumbbell diffuses slower than a single bead. However, choosing $V_0$ twice as large, $V_0=4$, we find a single bead diffusion with $D/D_0=1/\left(I_0(2)\right)^2=0.19...$ which is substantially slower than the corresponding stiff dumbbell diffusion. Here the elevator effect accelerates stiff dumbbells compared to single beads. Even a relatively soft dumbbell with $k=0.1$ diffuses slightly faster than a single bead. There is a critical value of $V_0$, $V_c=2.456...$, below which single beads diffuse always faster than dumbbells and above which stiff dumbbells are always faster. This value is found by requiring that the single bead's diffusion constant in a potential to equal $1/2$ which is the strong-spring limit of dumbbells with $x_\mathrm{eq}=a/2$, see Eq.~\ref{strong spr diff}. This leads to the condition $I_0\left(V_c/2\right)=\sqrt{2}$. In Fig.~\ref{elevator 2}, the diffusion constants of a dumbbell for this critical value of $V_0$ are also shown.

\begin{figure}
\includegraphics[scale=0.9]{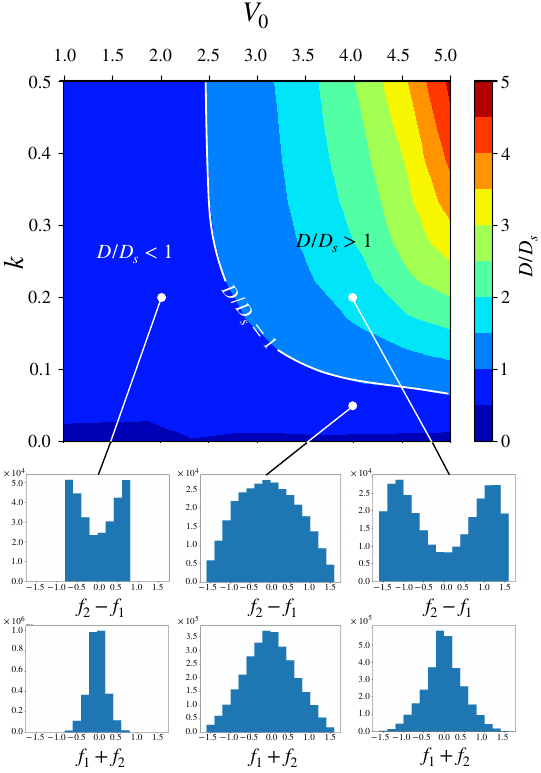}
\caption{\label{elevator contour} Top: Contour plot of the diffusion constant of a dumbbell with $x_\mathrm{eq}=a/2$ as a function of barrier height $V_0$ and spring constant $k$. The diffusion constant $D$ is rescaled by the theoretical single bead diffusion constant $D_s$, Eq.~\ref{bessel}, in the respective potential $V_0$. Therefore, levels above $1$ indicate that the dumbbell diffuses faster than the single bead. On the bottom, histograms of the difference of the forces, $f_2-f_1$, and their sum, $f_1+f_2$, acting on the beads of the dumbbell, extracted from a long single trajectory. We show here three combinations of barrier height $V_0$ and spring constant $k$: $V_0=2.0$, $k=0.2$ (left), $V_0=4.0$, $k=0.05$ (middle) and $V_0=4.0$, $k=0.2$ (right). To create a smooth contour plot, we performed a spline interpolation on a finite set of simulation data, namely all combinations of $V_0=1.0, 1.5,...,5.0$ and $k=0.0, 0.1,..., 0.5$.}
\end{figure}

Figure \ref{elevator contour} presents a comparison between single bead and dumbbell diffusion in the $\left(k, V_0\right)$-plane for $x_\mathrm{eq}=a/2$. Specifically, we present a contour plot of the ratio of the dumbbell diffusion constant, $D$, and the single bead diffusion constant, $D_s$. As can be seen, for sufficiently large values of $V_0$ and $k$ this ratio is always larger than one, $D/D_s>1$, i.e.~the diffusion of the dumbbell is then faster than that of a single bead. In order to get a more physical understanding of this speed-up, we created histograms of the difference and the sum of the forces on the two beads, $f_2-f_1$ and $f_1+f_2$, which are due to the underlying periodic potential experienced by the particles. Below the contour plot in Fig.~\ref{elevator contour}, we present histograms for three conditions. The histograms collect instances of these quantities along a sample trajectory. The plots in the middle presents the distributions for a soft spring, $k=0.05$, in a potential with $V_0=4.0$. The distribution of the differences, $f_2-f_1$, is broad and peaks at zero, the one of the sum, $f_2+f_1$, is a bit sharper. Increasing the stiffness fourfold has a dramatic effect on the distributions, see histograms on the right. In this case, the $f_2-f_1$-histogram covers the same range of forces but now two pronounced peaks appear at large negative and at large positive forces. This reflects two configurations: one where the dumbbell has a potential hill in between (beads pushed apart) and one where it has a valley in between (beads pushed together). However, despite the large forces acting on the beads, the histogram of the sum of the beads is rather sharp, showing that the forces on the two beads compensate each other. As a result, the dumbbell feels a rather flat landscape, reflecting the elevator effect. The same is the case for the histograms on the left hand side ($k=0.2$, $V_0=2.0$). However, the range of forces is smaller here, which reflects the smaller amplitude of the external potential. Even though the elevator effect plays a role in this case, the single bead still diffuses faster.

Finally we note that the diffusion of a dumbbell in a spatially periodic \textit{two-dimensional} potential was studied numerically in Ref.~\cite{zimmermann} where it was noted that the diffusion is enhanced when $a\approx 3x_\mathrm{eq}/2$ (a relation different from the one-dimensional case), see Fig.~4 in that paper. However, because of the orientational degree of freedom of the dumbbell in two dimensions, only a rather qualitative interpretation of this effect could be provided. Dumbbells show a clear-cut elevator effect only in a one-dimensional setting.

\section{Outlook}
We expect the results of this study to provide a foundation for exploring more complex systems, such as multi-dumbbell configurations or higher-dimensional periodic potentials. Longer bead-spring chains in periodic potentials lead to the so-called Frenkel-Kontorova model which was originally developed to explain dislocations in crystals \cite{Frenkel:431595, Kontorova:431596, Schiessel_Oshanin}. It has since been applied to a broad spectrum of physical systems, including charge-density waves \cite{floria1996dissipative}, and the motions of vortices in superconductors and Josephson junctions \cite{mclaughlin1978perturbation, watanabe1996dynamics, pedersen}. The Frenkel-Kontorova model can also be used to describe the diffusion along the DNA of the most abundant DNA-protein complex in eukaryotic cells, a DNA-wrapped protein cylinder called nucleosome \cite{Schiessel2023}. This type of diffusion is known to be caused by twist defects inside the wrapped DNA \cite{Lequieu2017, Niina2017, Rudnizky2019}. This process was first modeled by mapping nucleosomal DNA on a generalized Frenkel-Kontorova system, where the base pairs were treated as discrete units, namely a series of particles connected by elastic springs, reflecting the elasticity of the DNA molecule, see Ref.~\cite{kulic2003chromatin} and its active matter extension, Ref.~\cite{Klempahn24}. Due to constraints imposed by binding sites between the DNA and the protein core, nucleosomal DNA moves in a corkscrew manner. Since the intrinsic geometry of the base pair steps is sequence-dependent, different steps feel periodic bending energies with different amplitudes and phases as a function of their translational position. This even allows, in principle, for different base pair steps compensate each others contributions. This can lead to a different type of elevator effect, which can occur even when the periodicities of the bead-spring chain and of the potential are commensurate.

\acknowledgments
BAK was supported by the project ``Packaging and accessing DNA molecules'' with file number OCENW.KLEIN.089 of the research programme Open Competition ENW-KLEIN which is financed by the Dutch Research Council (NWO). HS was supported by the Deutsche Forschungsgemeinschaft (DFG, German Research Foundation) under Germany’s Excellence Strategy - EXC-2068 - 390729961.

\bibliography{apssamp}

\end{document}